\documentclass[conference, letterpaper]{IEEEtran}
\ifCLASSINFOpdf
  \usepackage[pdftex]{graphicx}
\else
\fi
\usepackage{amsmath}
\usepackage{amssymb}
\usepackage{amsfonts}
\usepackage{subfig}

\usepackage{color, colortbl}
\definecolor{Gray}{gray}{0.9}

\begin{document}
%
\title{Improving Broadcast Channel Rate Using Hierarchical Modulation}


\author{\IEEEauthorblockN{Hugo M{\'e}ric\IEEEauthorrefmark{1}\IEEEauthorrefmark{2},
J{\'e}r{\^o}me Lacan\IEEEauthorrefmark{2}\IEEEauthorrefmark{1},
Fabrice Arnal\IEEEauthorrefmark{3}, 
Guy Lesthievent\IEEEauthorrefmark{4} and
Marie-Laure Boucheret\IEEEauthorrefmark{2}\IEEEauthorrefmark{1}}
\IEEEauthorblockA{\IEEEauthorrefmark{1}T{\'e}SA, Toulouse, France}
\IEEEauthorblockA{\IEEEauthorrefmark{2}Universit{\'e} de Toulouse, Toulouse, France}
\IEEEauthorblockA{\IEEEauthorrefmark{3}Thales Alenia Space, Toulouse, France}
\IEEEauthorblockA{\IEEEauthorrefmark{4}CNES, Toulouse, France\\
Email: hugo.meric@isae.fr, jerome.lacan@isae.fr, fabrice.arnal@thalesaleniaspace.com,\\
guy.lesthievent@cnes.fr, marie-laure.boucheret@enseeiht.fr}}

\IEEEoverridecommandlockouts




\maketitle

\begin{abstract}

We investigate the design of a broadcast system where the aim is to maximise the throughput. This task is usually challenging due to the channel variability. Modern satellite communications systems such as DVB-SH and DVB-S2 mainly rely on time sharing strategy to optimize throughput. They consider hierarchical modulation but only for unequal error protection or backward compatibility purposes. We propose in this article to combine time sharing and hierarchical modulation together and show how this scheme can improve the performance in terms of available rate. We present the gain on a simple channel modeling the broadcasting area of a satellite. Our work is applied to the DVB-SH standard, which considers hierarchical modulation as an optional feature.
\end{abstract}


%
\IEEEpeerreviewmaketitle

\section{Introduction}

\begin{figure}[!ht]
\centering
\includegraphics[width = 0.85\columnwidth]{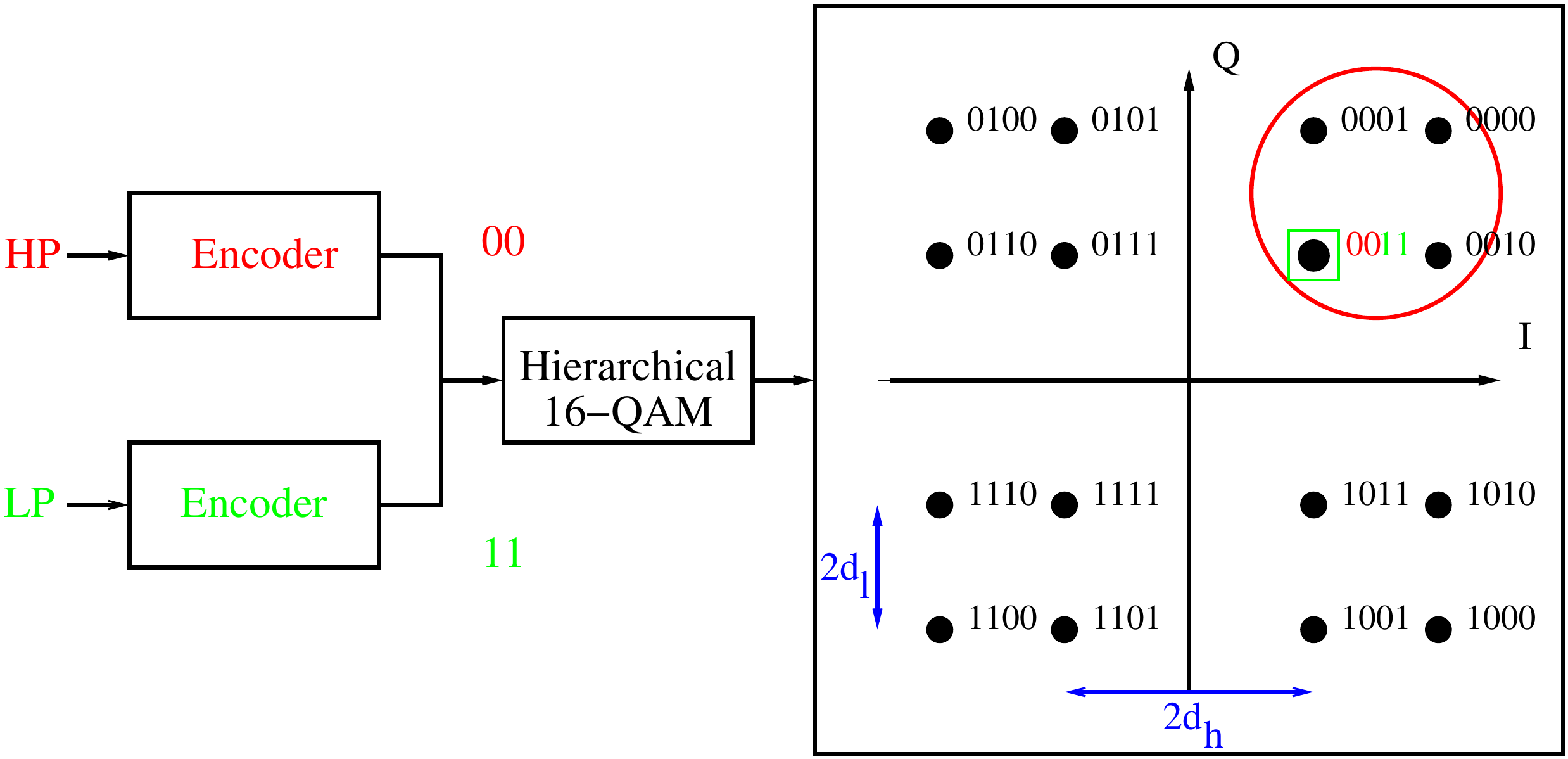}
\caption{Hierarchical Modulation using a non-uniform 16-QAM}
\label{hm_principle}
\end{figure}

In most broadcast applications, the signal-to-noise ratio (SNR) experienced by each receiver varies greatly. For instance, in satellite communications the channel quality decreases with the presence of clouds in Ku or Ka band, or with shadowing effects of the environment in lower bands. This leads to many difficulties to design an efficient (in terms of throughput) broadcast system. The first solution for broadcasting was to design the system for the worst-case reception, but this leads to poor performance as many receivers do not exploit their full potential. Then, two other schemes have been proposed in \cite{cover} and \cite{bergmans}: time division multiplexing with variable coding and modulation, and superposition coding. Time division multiplexing, or time sharing, allocates to each receiver a fraction of time where it can use the full channel with any modulation and error protection level. This solution is the most used in practice in standards today. In superposition coding, the available energy is shared among several data streams which are sent simultaneously in the same band. This scheme was introduced by Cover in \cite{cover} in order to improve the transmission rate from a single source to multiple receivers. When communicating with two receivers, the principle is to superimpose information for the receiver with the best SNR. This superposition can be done directly at the forward error correction (FEC) encoding level or at the modulation level. Hierarchical modulation is a practical implementation of superposition coding. Figure~\ref{hm_principle} presents the principle of hierarchical modulation with a non-uniform 16-QAM and the mapping used in this paper. The idea is to merge two different streams at the modulation step. The High Priority (HP) stream is used to select the quadrant, and the Low Priority (LP) stream selects the position inside the quadrant. The HP stream is dedicated to receivers with bad channel quality, unlike the LP stream which requires a good SNR to be decoded. Today, hierarchical modulation is used mainly for unequal error protection. For instance in \cite{svc_hm}, SVC encoded video \cite{svc} is protected using hierarchical modulation. The base layer of the video is transmitted in the HP stream, while the enhanced layer is carried by the LP stream. Another usage is backward compatibility \cite{backward_compatibility}. The DVB-S2 standard is called upon to replace DVB-S, but many DVB-S receivers are already installed. Then the hierarchical modulation helps to the migration by allowing the DVB-S receivers to operate. In \cite{local_content} , the authors propose to provide local content with hierarchical modulation. The principle is to carry local information that is of interest to a particular geographic area in the LP stream, while the HP stream transmits global content. Finally, other works  improve the performance of relay communication system \cite{relaycom} or OFDMA-based networks \cite{icc10}. 

Our work focuses on using hierarchical modulation in modern broadcast systems to increase the transmission rate. In satellite standards such as DVB-SH or DVB-S2, time sharing and hierarchical modulation are not used simultaneously. This article investigates the performance, in terms of throughput, of a broadcast system where these two schemes work together. We focus on the DVB-SH standard and show on a simple example that the gain can be significant, up to 17\%. 
 
The paper is organized as follows: Section~\ref{achievable_rates} computes the throughput for the different broadcasting solutions, Section~\ref{application} studies the performance of each scheme and Section~\ref{conclusion} concludes the paper by summarizing the results.

\section{Throughput computation}\label{achievable_rates}

This section introduces the throughput for the two following schemes:
\begin{itemize}
\item Time sharing strategy with no hierarchical modulation, referred as \emph{classical time sharing}.
\item Time sharing strategy where hierarchical modulation is allowed, referred as \emph{hierarchical modulation time sharing}.
\end{itemize}

\subsection{Classical time sharing}

We consider a broadcast system with $n$ receivers, each one with a particular signal-to-noise ratio $SNR_i$ ($i=1,..,n$). Given $SNR_i$, receiver $i$ has a rate $R_i$ which corresponds to the best rate it can afford. This rate is the amount of useful data transmitted on the link and depends on the modulations and coding rates available in the system. For instance, if the modulation is a QPSK and the coding rate is 1/3, then the rate equals $2 \times 1/3$ bit/symbol. In our study, the physical layer is based on the DVB-SH standard \cite{sh, DVBSH}. Table~\ref{cts} resumes the modulations and coding rates used for the classical time sharing scheme.
\begin{table}[!ht]
\renewcommand{\arraystretch}{1.1}
\caption{Classical time sharing physical layer}
\label{cts}
\centering
\begin{tabular}{c||c} 
\hline
Modulations & QPSK, 16-QAM \\ 
\hline
Coding rates & 1/5, 2/9, 1/4, 2/7, 1/3, 2/5, 1/2, 2/3\\
\hline 
\end{tabular}
\end{table}

The time sharing scheme allocates a fraction of time $t_i$ to receiver $i$. The performance, in terms of throughput, results from the time allocation. We define the average rate for receiver $i$ as $t_iR_i$. In our study, we are interested to offer the \emph{same average rate to everyone}. Then, we need to solve
\begin{equation}
\begin{cases}t_iR_i = t_jR_j \text{ } \forall i,j \\ \sum_{i} t_{i} = 1. \end{cases}
\label{sh_equation}
\end{equation} 

It is easy to verify that the set of $t_i$ defined as follows,
\begin{equation}
t_{i} = \frac{\prod_{k\neq i} R_k}{\sum_{j=1}^n \left(\prod_{k\neq j} R_k \right)},
\label{fraction_time}
\end{equation}

\noindent is the unique solution of (\ref{sh_equation}). We remark that increasing $R_i$ reduces $t_i$, which is a consequence of our rate policy. Finally, the average rate offered to each receiver is
\begin{equation}
R = \frac{\prod_{k} R_k}{\sum_{j=1}^n \left(\prod_{k\neq j} R_k \right)}.
\label{rate}
\end{equation}

\subsection{Hierarchical modulation time sharing}

We first give some definitions and then compute the throughput for the hierarchical modulation time sharing scheme. As mentioned before, hierarchical modulations merge several streams in a same symbol. They often use non-uniform constellation where the points are not  uniformly distributed in the plane. The geometry of non-uniform constellations is described using the constellation parameter. The DVB-SH standard considers the hierarchical 16-QAM presented in Figure~\ref{hm_principle}. The constellation parameter $\alpha$ is defined by $d_h/d_l$, where $2d_h$ is the minimum distance between two constellation points carrying different HP bits and $2d_l$ is the minimum distance between any constellation point. Typically, we have $\alpha \ge 1$, where $\alpha=1$ corresponds to the uniform 16-QAM, but it is also possible to have $\alpha \le 1$. At a given energy per symbol ($E_s$), when $\alpha$ grows, the constellation points in each quadrant become farther to the I and Q axes. Thus it is easier to decode the HP stream. However, in the same quadrant, the points get closer and the LP stream requires a good channel quality to be decoded.

We define the QPSK parameter as the minimum distance between two points in a QPSK constellation. As shown in Figure~\ref{hm_principle}, the hierarchical 16-QAM is the superposition of two QPSK modulations, one with parameter $2(d_h+d_l)$ carrying the HP stream and the other with parameter $2d_l$ carrying the LP stream. Thus the energy ratio between the two streams is
\begin{equation}
\frac{E_{hp}}{E_{lp}} = (1+\alpha)^2 ,
\end{equation}

\noindent where $E_{hp}$ and $E_{lp}$ corresponds to the energy allocated to the HP and LP streams respectively \cite{local_content}. The standard \cite{sh} recommends two values for $\alpha$: 2 and 4. In fact, DVB-SH also considers $\alpha=1$ but only for the Variable Coding Modulation mode. The values 2 and 4 are defined in order to provide unequal protection. From an energy point of view, this amounts to give 90\% ($\alpha=2$) or 96\% ($\alpha=4$) of the available energy to the HP stream. However, our goal is not to provide unequal protection. It is why new values of $\alpha$ are used in order to boost the throughput of the system. Our simulations include $\alpha=1$, $\alpha=0.8$ and $\alpha=0.5$, which provide a better repartition of the energy: the HP stream contains 80\%, 76\% and 69\% of the total power. 

\begin{figure}[!ht]
\centering
\includegraphics[width = 0.8\columnwidth]{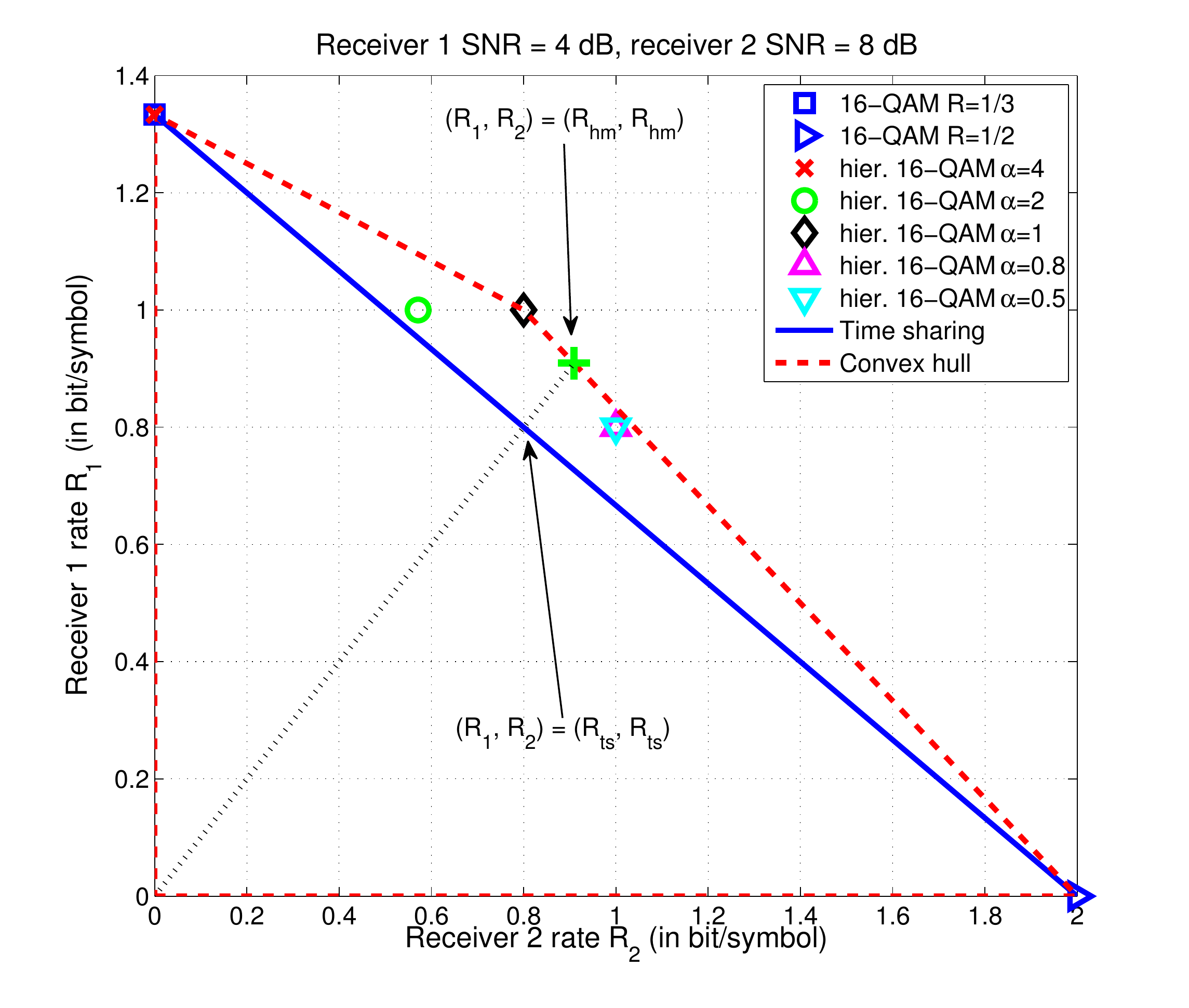}
\caption{Achievable rate region}
\label{sh_cr_4_8}
\end{figure}

To determine the throughput of this scheme, we first investigate the case with \emph{two receivers}. We begin by computing the rates offered by all the possible modulations (Table~\ref{cts}), including the hierarchical 16-QAM considered in the DVB-SH standard. Thus we obtain a set of achievable $(R_1, R_2)$ rate pairs, where $R_i$ is the rate of receiver $i$, $i=1,2$. Classical modulations (Table~\ref{cts}) achieve rate pairs of the form $(R_1,0)$ and $(0,R_2)$, while the hierarchical 16-QAM allows rates pairs of the form $(R_1, R_2)$. When the hierarchical modulation is used, we assume that \emph{the LP stream is dedicated to the the receiver with the largest SNR}. Moreover, when two sets of rates $(R_1,R_2)$ and $\left(R_1^*,R_2^*\right)$ are achievable, the time sharing achieves any rate pair 
\begin{equation}
\scriptstyle
\tau (R_1,R_2) + (1-\tau)(R_1^*,R_2^*) = (\tau R_1+(1-\tau)R_1^* , \tau R_2+(1-\tau)R_2^* ) ,
\end{equation}

\noindent where $0 \le \tau \le 1$ is the proportion of time allocated to $(R_1,R_2)$. The achievable rate region is the convex hull of the set of achievable $(R_1,R_2)$ rate pairs. Figure~\ref{sh_cr_4_8} presents the achievable rate region for two receivers, one with a SNR of 4 dB and the other 8 dB. We also represent the classical time sharing achievable rate region. Note that we only represent the rate pairs which are relevant. The points for $\alpha=4$ and $\alpha=2$ are given in \cite[Table 7.11]{sh} where 0.3 dB needs to be removed due to the pilots. The other points (i.e., $\alpha=1$, $\alpha=0.8$ and $\alpha=0.5$) are computed using the method described in \cite{wts}. All the decoding thresholds are given in Table~\ref{decoding_threshold} and the Appendix details the computation for non standard points on one example. As we are interested to offer the \emph{same rate} to the receivers, we calculate the intersection of the convex hull with the curve $y=x$, which corresponds to the cross in Figure~\ref{sh_cr_4_8}. The interest of using hierarchical modulation is clear as the final rate results of time sharing between two operating points where one is obtained using the hierarchical 16-QAM with $\alpha=1$.

Figure~\ref{sh_cr_4_8} deserves few more comments. First, the rates obtained for the classical time sharing strategy come from the 1/3 16-QAM for the receiver at 4 dB and the 1/2 16-QAM for the other. As mentioned before, it is the best that each receiver can afford. Concerning $\alpha=4$, it leads to a zero rate for the receiver who decodes the LP stream. We have already said that the points in one quadrant get closer when $\alpha$ grows (at a given $E_s$), which requires the receiver decoding the LP stream to have a good channel quality as observed in Table~\ref{decoding_threshold}. This is why large values of $\alpha$ are interesting only when some receivers experience very good channel quality. Using $\alpha=4$ and the smallest coding rate in the standard (1/5), the decoding threshold is 10.3 dB and then the receiver with a SNR of 8 dB can not decode anything. We also remark that $\alpha=0.5$ and $\alpha=0.8$ give the same rate pairs, which is a consequence of the coding rate discretization. In fact, a close look to Table~\ref{decoding_threshold} (colored cells) shows the HP and LP streams use the $2/5$ and $1/2$  coding rates respectively for both $\alpha$. Finally, the new values of $\alpha$ give the best results in that case.

\begin{table}[!ht]
\renewcommand{\arraystretch}{1.1}
\setlength{\tabcolsep}{0.13cm}
\caption{Hierarchical 16-QAM decoding thresholds (in dB)}
\label{decoding_threshold}
\centering
\begin{tabular}{c||c|c|c|c|c|c|c|c|c|c} 
\hline
Coding & \multicolumn{2}{c|}{$\alpha=4$} & \multicolumn{2}{c|}{$\alpha=2$} & \multicolumn{2}{c|}{$\alpha=1$} &  \multicolumn{2}{c|}{$\alpha=0.8$} & \multicolumn{2}{c}{$\alpha=0.5$} \\ 
\cline{2-11}
rate & HP & LP & HP & LP & HP & LP & HP & LP & HP & LP \\
\hline
1/5 & -3.6 & 10.3 & -3.2 & 6.2 & -2.5 & 3.7 & -2.1 & 3.2 & -1.4 & 2.6 \\
\hline
2/9 & -3.1 & 10.8 & -2.6 & 6.8 & -1.9 & 4.1 & -1.6 & 3.6 & -0.8 & 2.9 \\
\hline 
1/4 & -2.5 & 11.4 & -2 & 7.3 & -1.2 & 4.6 & -0.9 & 4.1 & 0 & 3.4 \\
\hline 
2/7 & -1.8 & 12.1 & -1.3 & 8 & -0.4 & 5.2 & 0 & 4.7 & 0.9 & 3.9 \\
\hline 
1/3 & -0.9 & 12.9 & -0.4 & 8.8 & 0.7 & 6 & 1.1 & 5.4 & 2.2 & 4.6 \\
\hline 
2/5 & 0.2 & 14 & 2 & 6.8 & 9.9 & 6.9 & \cellcolor{Gray} 2.5 & 6.3 & \cellcolor{Gray} 3.8 & 5.4 \\
\hline 
1/2 & 1.6 & 15.3 & 7.3 & 11.3 & 3.7 & 8.1 & 4.4 & \cellcolor{Gray} 7.5 & 6.2 & \cellcolor{Gray} 6.5 \\
\hline 
2/3 & 3.9 & 17.5 & 4.9 & 13.4 & 7 & 10.2 & 8.1 & 9.5 & 11 & 8.4 \\
\hline   
\end{tabular}
\end{table}

We now consider a broadcast system with $n$ receivers. The first step is to group the receivers in pairs in order to use the hierarchical modulation. Then for each pair, the achievable rate is computed as described previously. Finally, we need to equalize the rate between each receiver. This is done by time sharing. The rate is known for each receiver and (\ref{fraction_time}) can be applied. For instance, consider a system where a receiver with a SNR of 4 dB is in pair with a receiver having a SNR of 8 dB. The rate for each receiver is obtained using the hierarchical 16-QAM ($\alpha=1$) a fraction of time $a_1$ and the 16-QAM a fraction of time $a_2$ as shown in Figure~\ref{sh_cr_4_8}. When equalizing the rates, (\ref{fraction_time}) gives the same fraction of time $t$ to both receivers. At last, the broadcast system allocates to our pair of receivers the hierarchical 16-QAM ($\alpha=1$) for a time proportion $2t \times a_1$ and the 16-QAM for $2t \times a_2$. 

Finally, the performance obtained here is a consequence of our \emph{rate policy}: we choose to offer the same rate to everyone. However, our solution can be adapted to any other policy. For instance, if the system contains premium users who pay to get a better rate, the fraction of time dedicated to these receivers will be computed according to this new policy.

\section{Application to broadcast channel systems}\label{application}

This section compares the schemes described in Section~\ref{achievable_rates} on two broadcast channels. The first one consists of one source and two receivers, which is the simplest broadcast channel. Then, the performance is evaluated on an advanced example with one transmitter and six receivers.

\subsection{Channel with two receivers}

\begin{figure}[!ht]
\centering
\includegraphics[width = 0.83\columnwidth]{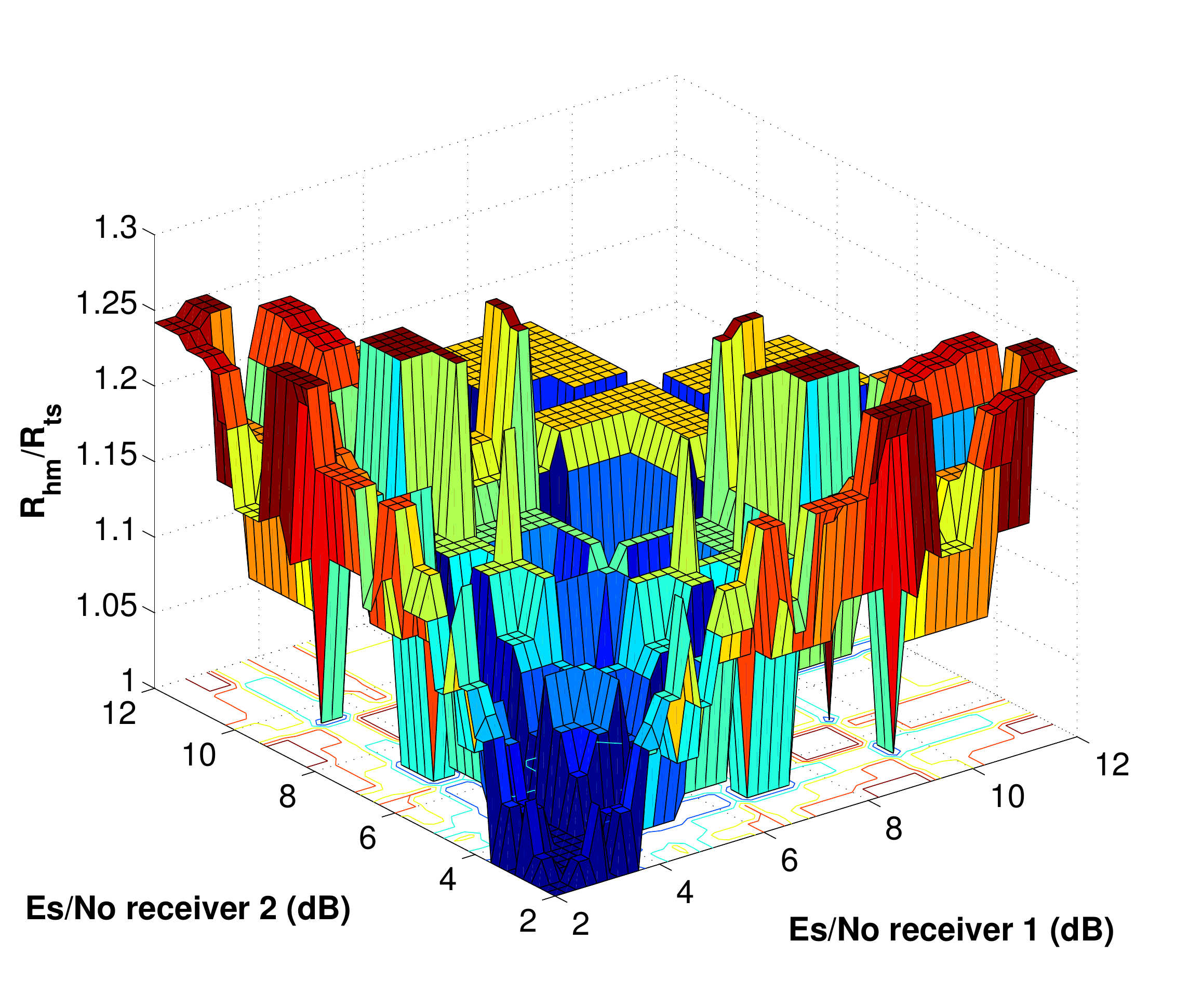}
\caption{Throughput ratio according to receivers' SNR}
\label{sh_gain}
\end{figure}

\begin{figure}[!ht]
\centering
\subfloat[Receiver 1 SNR: 2 dB]{\includegraphics[width=0.44\columnwidth]{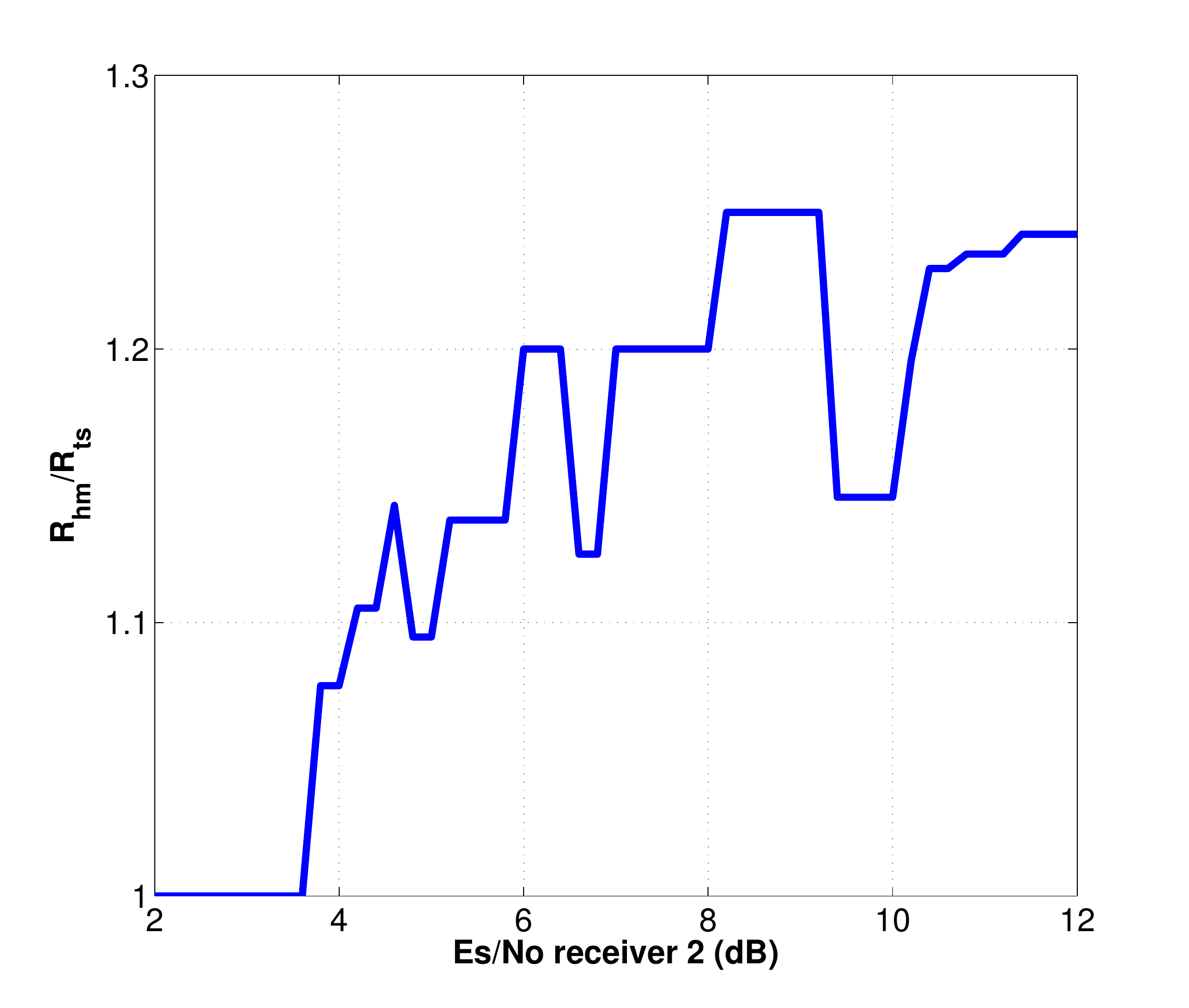}%
\label{gain2D_2}}%
\hfil
\subfloat[Receiver 1 SNR: 4 dB]{\includegraphics[width=0.44\columnwidth]{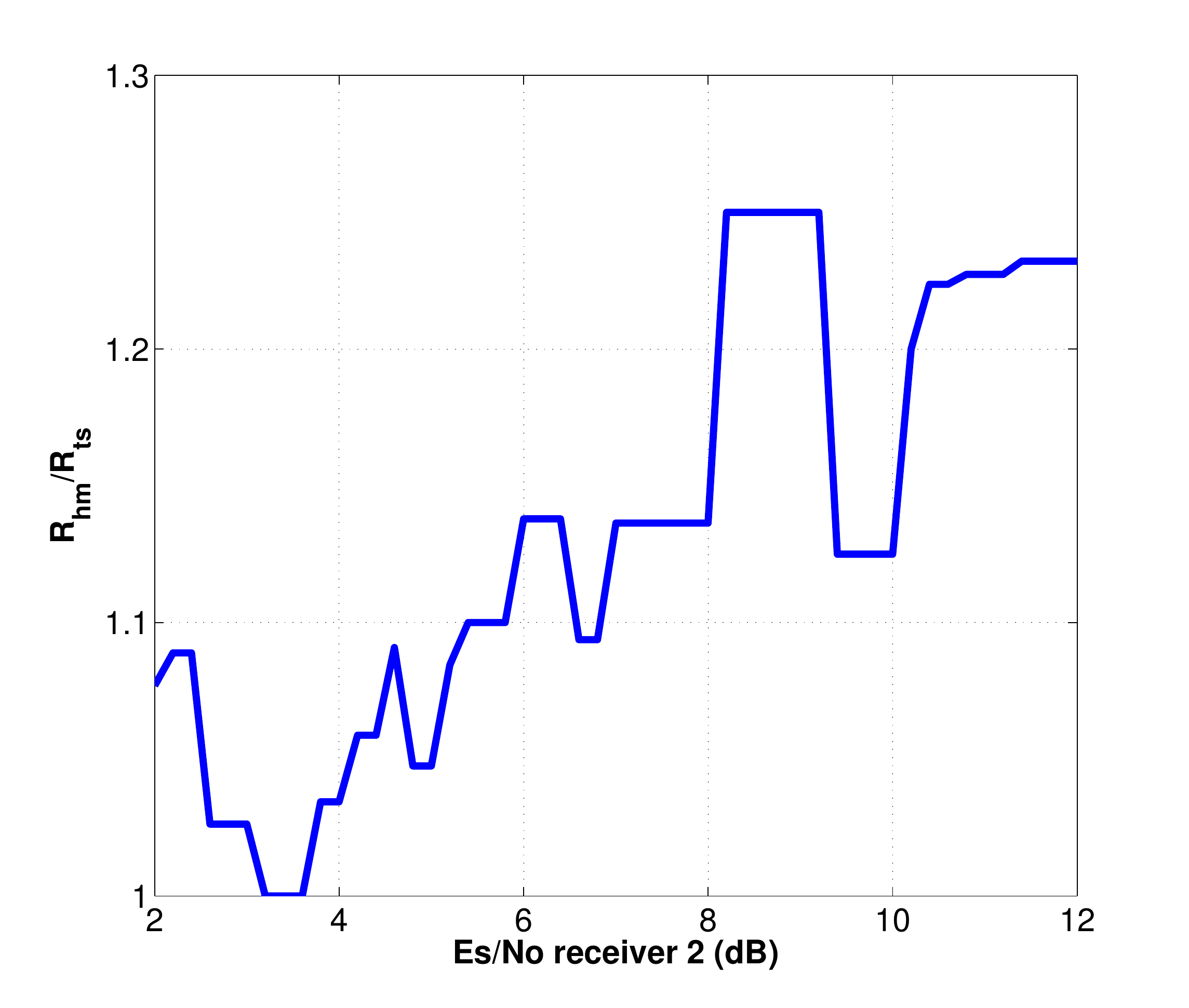}%
\label{gain2D_4}}

\subfloat[Receiver 1 SNR: 6 dB]{\includegraphics[width=0.44\columnwidth]{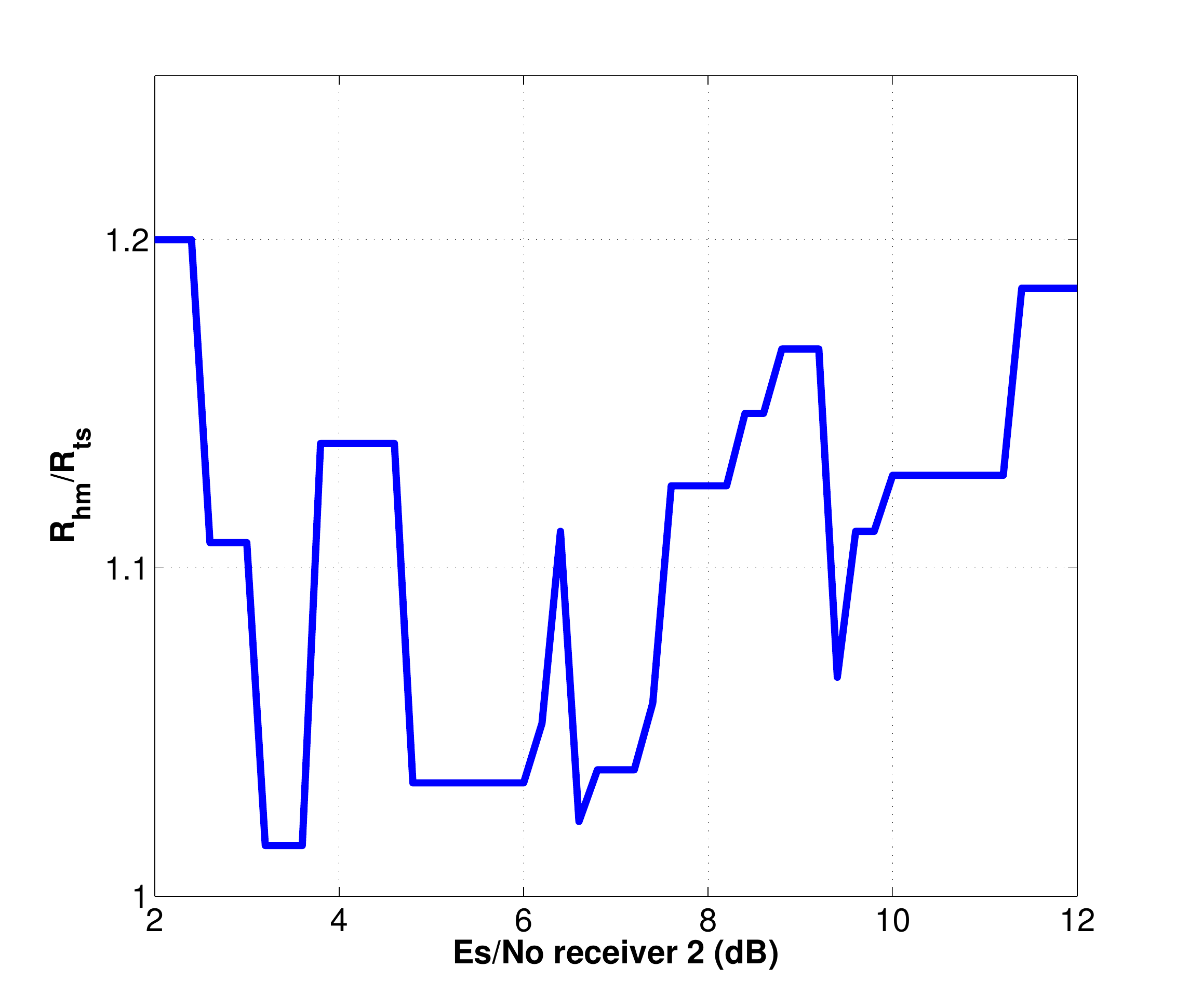}%
\label{gain2D_6}}%
\hfil
\subfloat[Receiver 1 SNR: 8 dB]{\includegraphics[width=0.44\columnwidth]{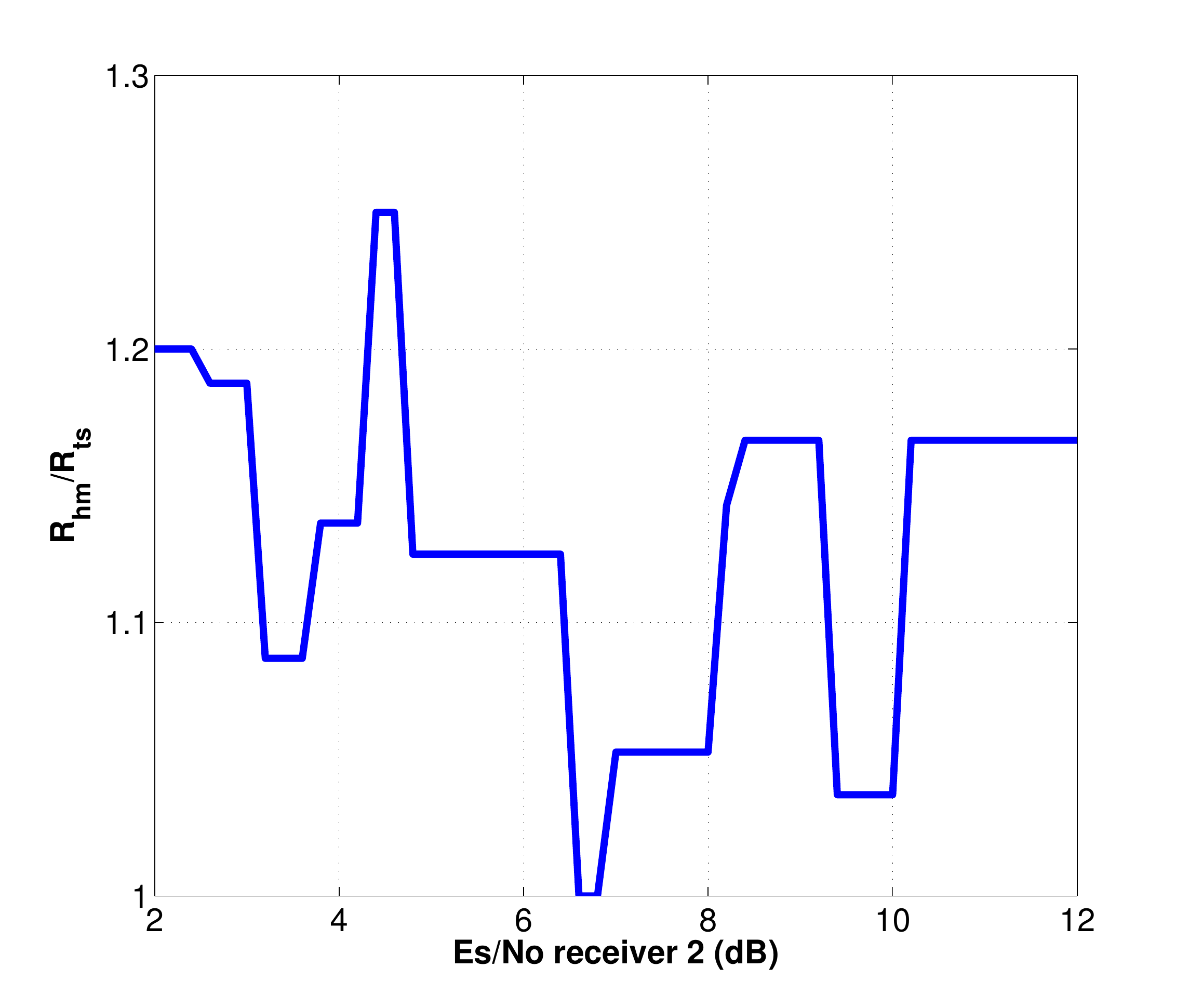}%
\label{gain2D_8}}

\subfloat[Receiver 1 SNR: 10 dB]{\includegraphics[width=0.44\columnwidth]{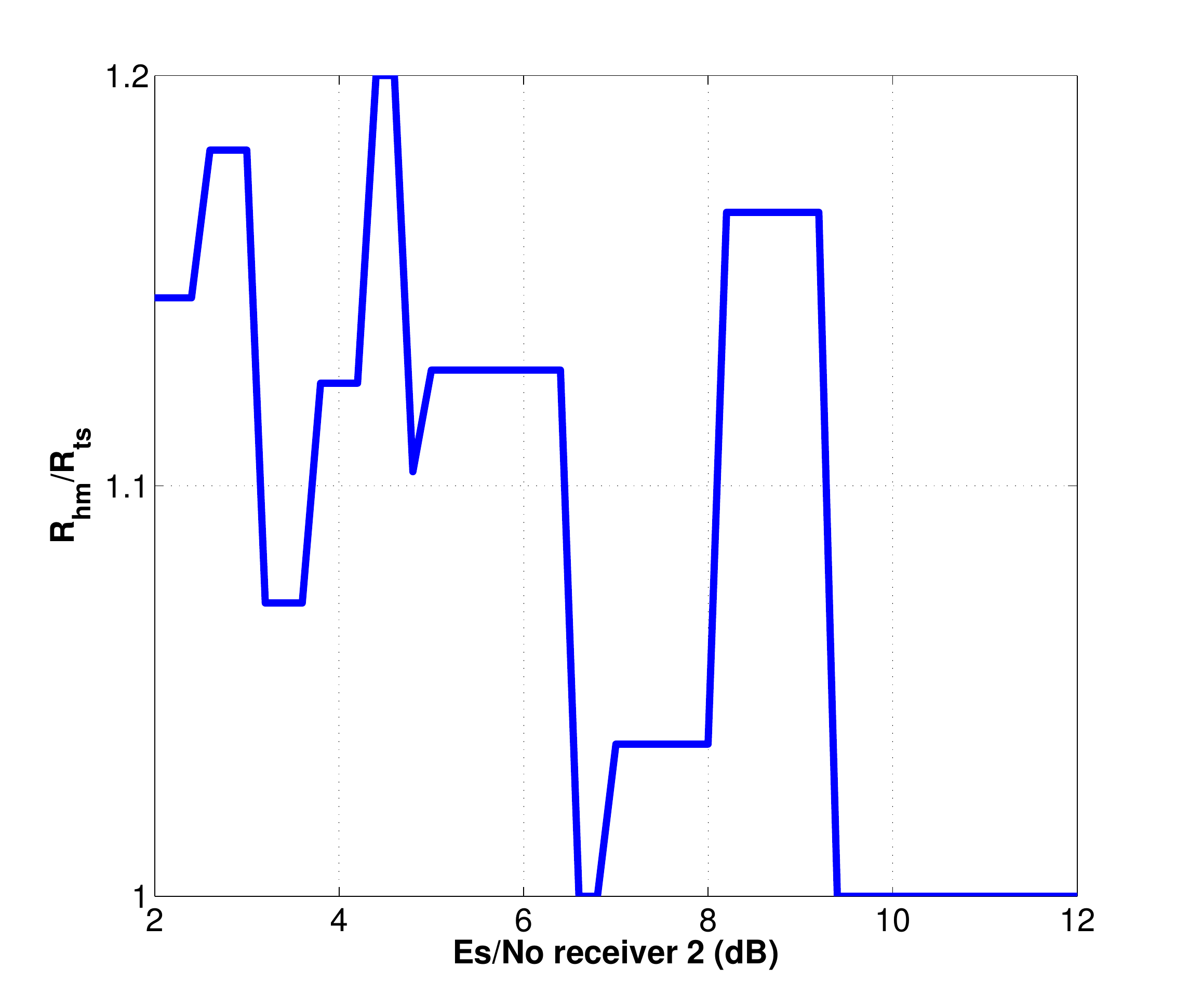}%
\label{gain2D_10}}%
\hfil
\subfloat[Receiver 1 SNR: 12 dB]{\includegraphics[width=0.44\columnwidth]{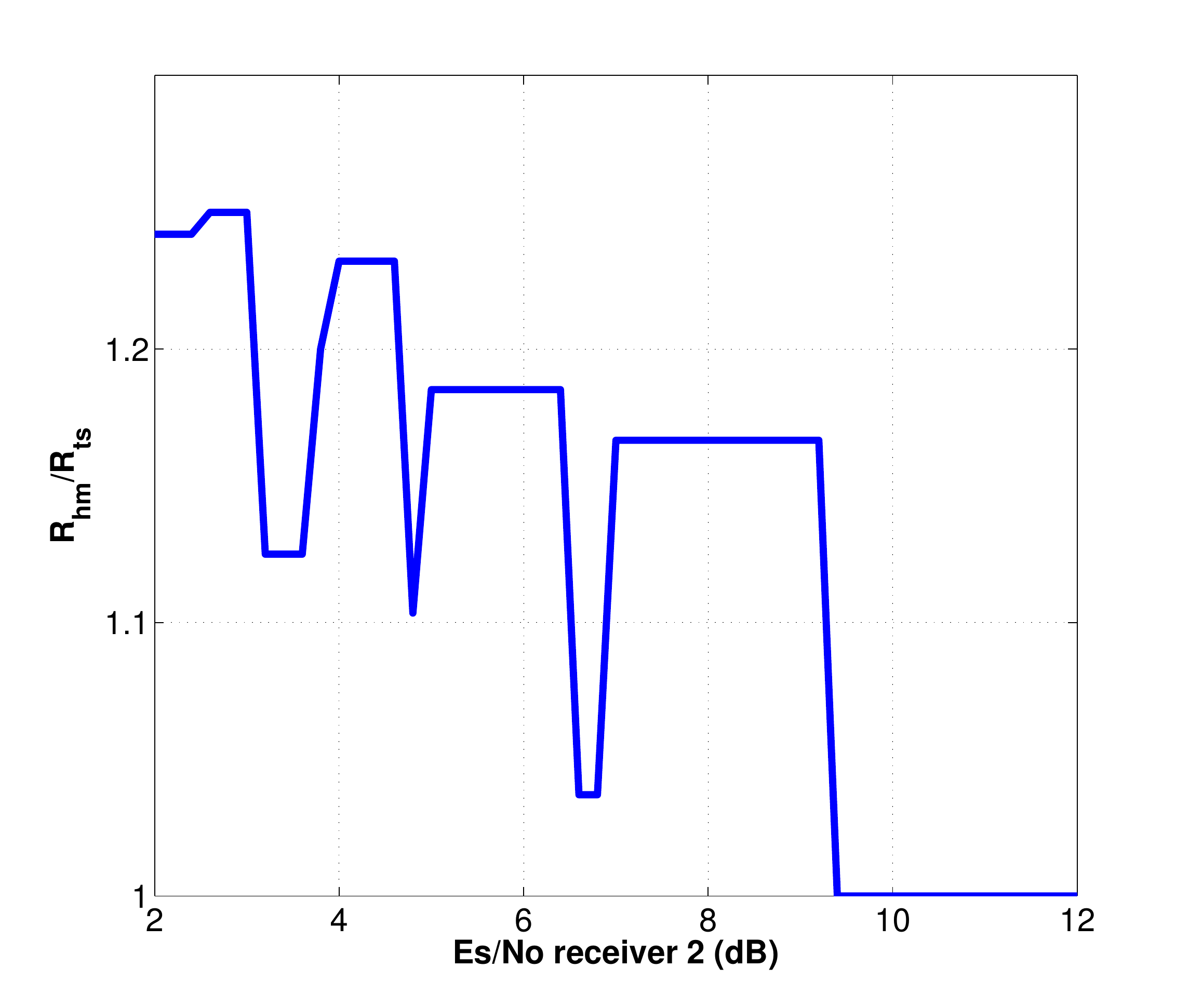}%
\label{gain2D_12}}

\caption{Throughput ratio}
\label{gain2D}
\end{figure}

We consider one source communicating with two receivers. Figure~\ref{sh_cr_4_8} already shows the throughput gain of the combined scheme in comparison to the classical time sharing strategy when one receiver has a SNR of 4 dB and the other a SNR of 8 dB. Figure~\ref{sh_gain} presents the throughput ratio $R_{hm}/R_{ts}$ against the receivers' SNR, where $R_{ts}$ and $R_{hm}$ are the throughput offered by the classical and hierarchical modulation time sharing respectively. It allows to have a global idea on the performance. A more detailed study can be done with Figure~\ref{gain2D}, which corresponds to cuts in Figure~\ref{sh_gain} for various SNR values of the receiver 1.

In general, the gains are more important when the SNR difference between the two receivers is large. For instance, in Figure~\ref{gain2D_2} and Figure~\ref{gain2D_4}, when receiver 1 experiences a bad channel quality, if the SNR of receiver 2 is large, the gain is almost 25\%. Moreover, in all figures, a gain of at least 20\% is observed. We also remark there is no gain when SNR is both small (e.g., Figure~\ref{gain2D_2}) or large (e.g., Figure~\ref{gain2D_10} and Figure~\ref{gain2D_12}). The first comment is helpful when we have to group the receivers in pairs in a large broadcast system. In fact, the next example will show that the grouping strategy has a great impact on the performance of the system, and then must be chosen carefully. Finally, the variations of all the curves in Figure~\ref{gain2D} can be explained as we compute the ratio of two increasing staircase functions, $R_{hm}$ and $R_{ts}$.

\subsection{Advanced example}

We consider one source communicating with six receivers ($Rec_i, i=1..6$). The channel is intended to model a satellite spot beam with clear sky conditions. At the center of the beam, the channel quality is good and there are few receivers, while away from the center, the SNR decreases but the number of receivers increases. Two parameters describe the channel: $SNR_{max}$ is the SNR at the center of the beam and $\Delta$ is used to characterize the attenuation due to the distance from the center of the beam. We assume that the SNR distribution is as follows:   
\begin{itemize}
\item $Rec_1$ has a SNR of $SNR_{max}-\Delta$
\item $Rec_2$ and $Rec_3$ have a SNR of $SNR_{max}-2\Delta$
\item $Rec_4$, $Rec_5$ and $Rec_6$ have a SNR of $SNR_{max}-3\Delta$
\end{itemize}

\begin{figure}[!ht]
\centerline{
\subfloat[Strategy A]{\includegraphics[width=0.32\columnwidth]{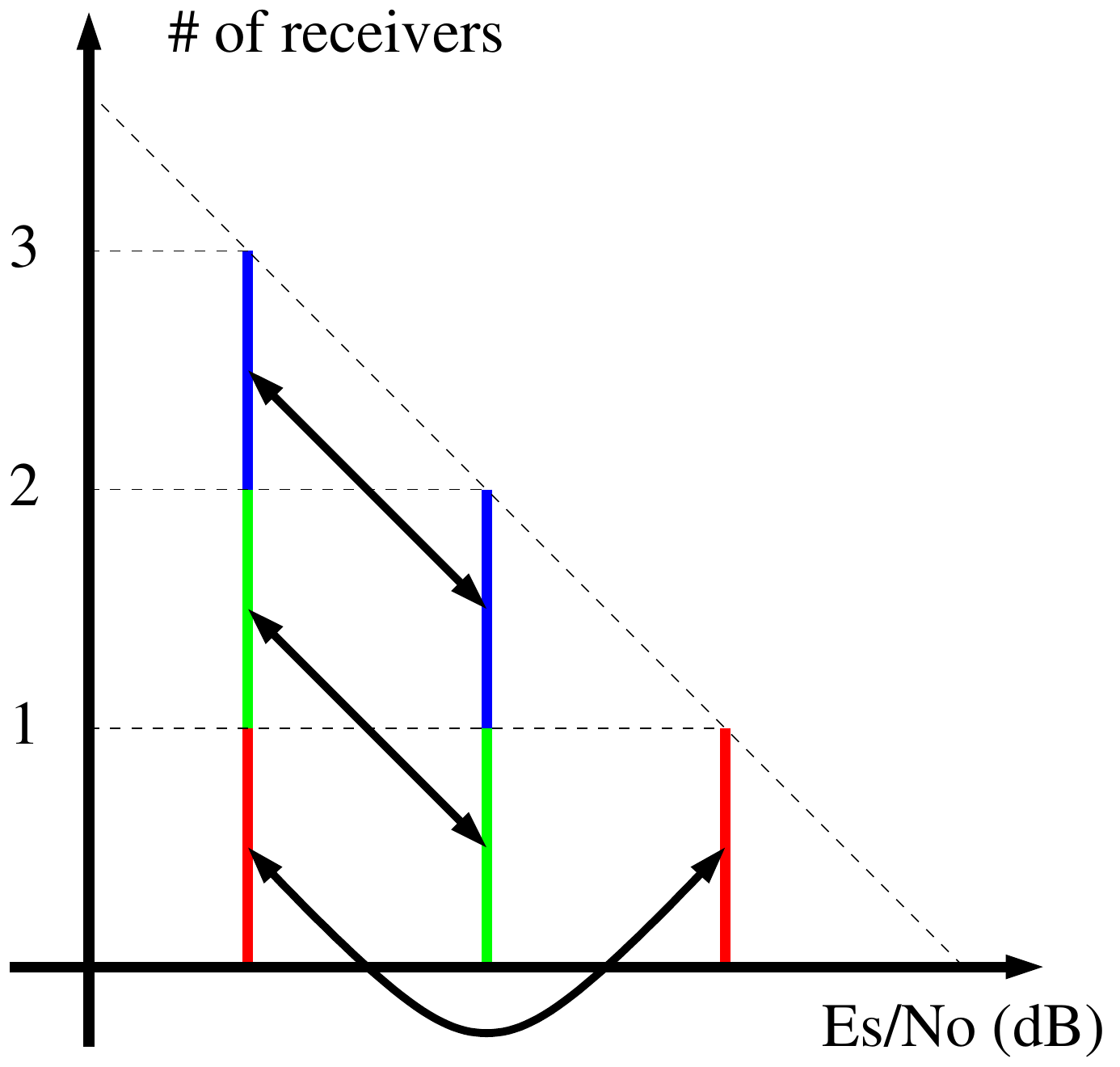}%
\label{grouping_1}}%
\hfil
\subfloat[Strategy B]{\includegraphics[width=0.32\columnwidth]{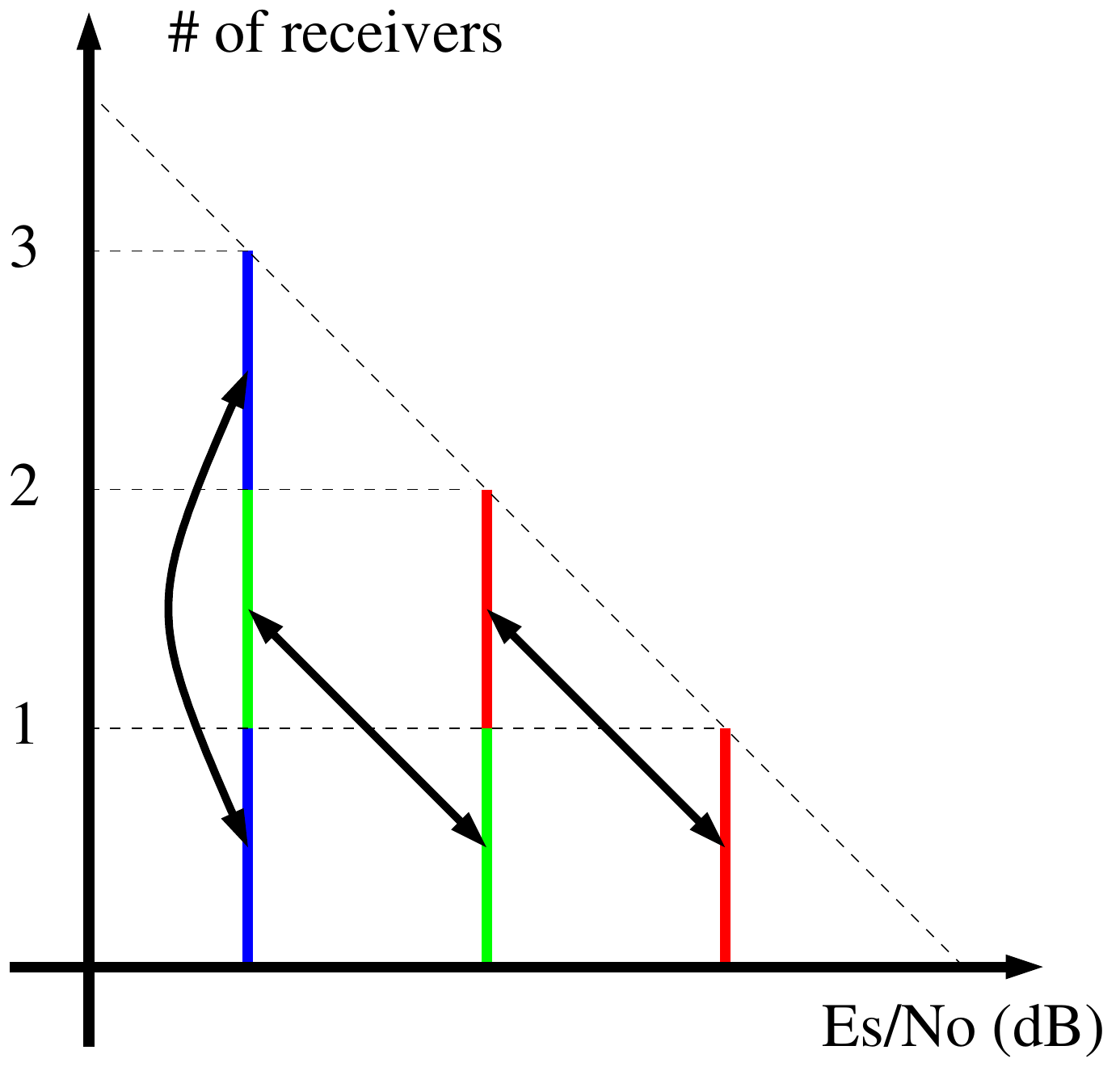}%
\label{grouping_2}}
\hfil
\subfloat[Strategy C]{\includegraphics[width=0.32\columnwidth]{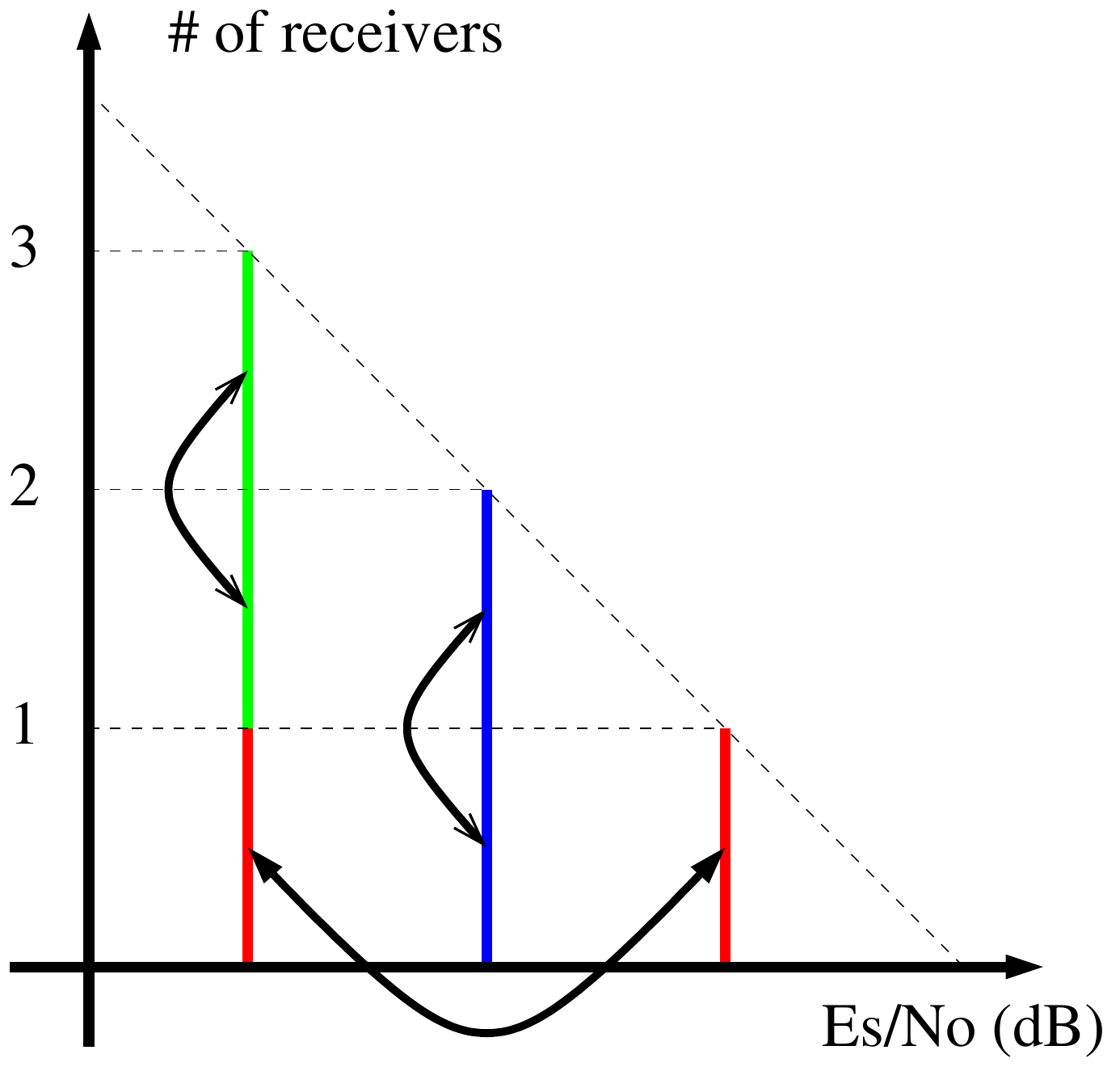}%
\label{grouping_3}}
}%
\caption{Grouping strategies}
\label{grouping_strategies}
\end{figure}

Using the DVB-SH guidelines \cite{sh} and (\ref{rate}), the classical time sharing strategy is easily treated. For the hierarchical modulation time sharing, we first need to group the receivers in pairs. For instance, we can group  $Rec_1$ with $Rec_4$, $Rec_2$ with $Rec_5$ and $Rec_3$ with $Rec_6$. In fact, the performance only depends on the SNRs of the receivers in pairs. Then the previous grouping is equivalent in terms of performance to the following scheme: $Rec_1$ with $Rec_6$, $Rec_2$ with $Rec_4$ and $Rec_3$ with $Rec_5$. These two groupings refer to strategy A in Figure~\ref{grouping_strategies}. Thus only three grouping strategies have to be considered. Figure~\ref{grouping_strategies} presents the three strategies studied for this example. 

\begin{figure}[!ht]
\centering
\includegraphics[width = 0.8\columnwidth]{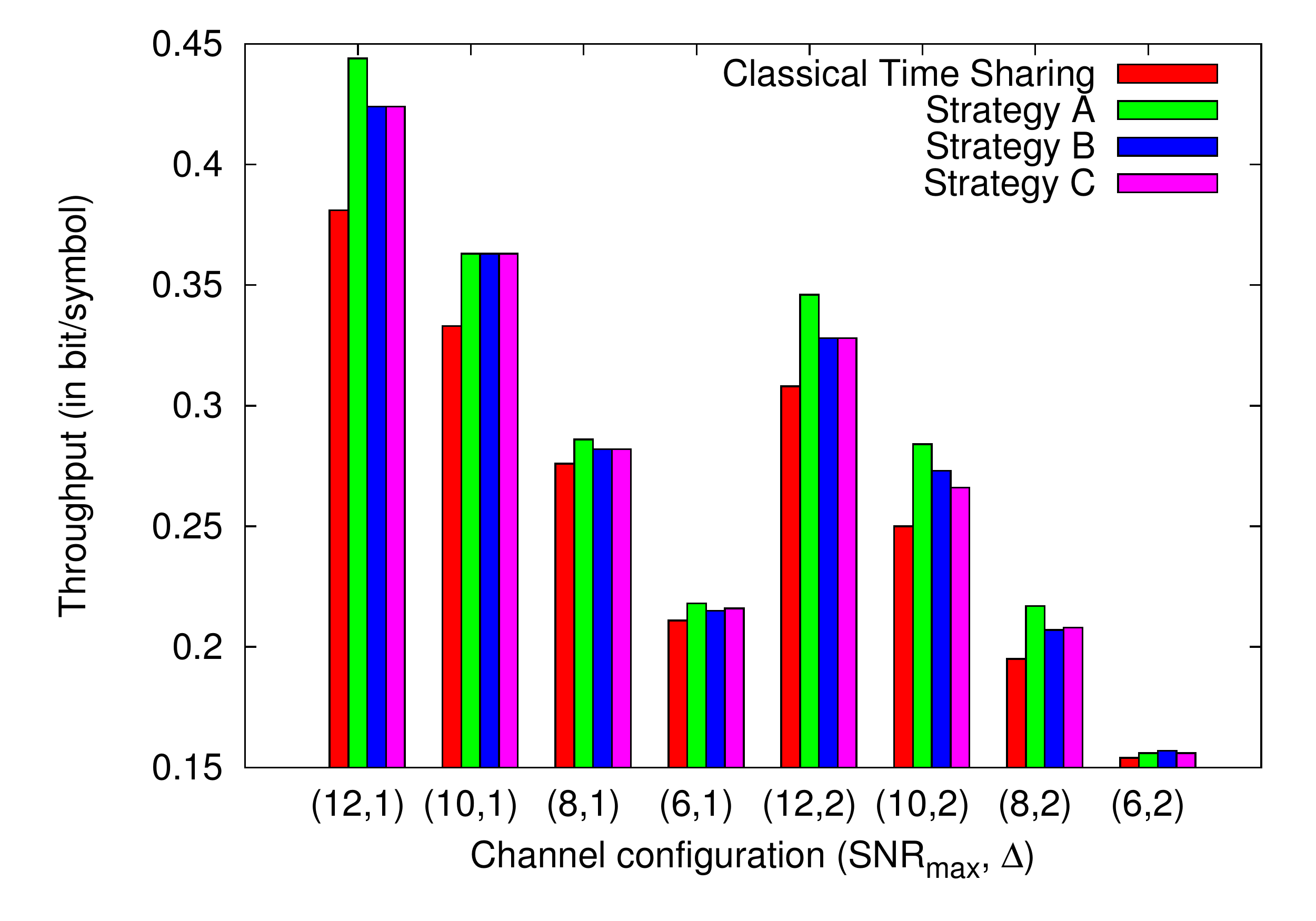}
\caption{Throughput according to channel configuration}
\label{throughput_bc}
\end{figure}

For each grouping strategy, we compute the best throughput available for the receivers when they implement the classical or hierarchical modulation time sharing. We illustrate its computation on an example. We consider the classical time sharing strategy with the channel parameters $(SNR_{max},\Delta)=(10\text{ dB},2\text{ dB})$. $Rec_1$ has a SNR of 8 dB and can decode the 1/2 16-QAM (best possible rate), $Rec_2$ and $Rec_3$ have a SNR of 6 dB and can decode the 2/5 16-QAM and finally $Rec_4$, $Rec_5$ and $Rec_6$ have a SNR of 4 dB and can decode the 1/3 16-QAM (or 2/3 QPSK) \cite{sh}. The average rate is computed with (\ref{rate}). For the hierarchical modulation time sharing, the calculation is done the same way for each strategy. Figure~\ref{throughput_bc} give the throughputs for the different solutions. The abscissa corresponds to the channel parameters $(SNR_{max},\Delta)$ and the hierarchical modulation time sharing corresponds to the histograms labeled Strategy A, B and C.

Finally, Table~\ref{gain_sh} presents the throughput gain for several channel configurations. We obtain a maximum gain of 17\% with the strategy A and a large SNR channel. For all strategies, we have an improvement of the throughput in comparison to the classical time sharing strategy. However, the strategy A obtains the best results, sometimes two times better than the other strategies. It shows the grouping strategy greatly impacts on the final performance of the system. Moreover, strategy A has the largest average SNR difference between two receivers in pairs. This supports the observation made on the previous example and give an idea on how to group the receivers in a large broadcast system.

\begin{table}[!ht]
\renewcommand{\arraystretch}{1.1}
\caption{Throughput gain}
\label{gain_sh}
\centering
\begin{tabular}{c|c|c|c}
\hline
$(SNR_{max}, \Delta)$ & Strategy A & Strategy B & Strategy C \\
\hline
\rowcolor{Gray}
(12 dB, 1 dB) & 17\% & 11\% & 11\% \\
\hline
(10 dB, 1 dB) & 9\% & 9\% & 9\% \\
\hline
\rowcolor{Gray}
(8 dB, 1 dB) & 4\% & 2\% & 2\% \\
\hline
(6 dB, 1 dB) & 3\% & 2\% & 2\% \\
\hline
\rowcolor{Gray}
(12 dB, 2 dB) & 12\% & 6\% & 6\% \\
\hline
(10 dB, 2 dB) & 14\% & 9\% & 6\% \\
\hline
\rowcolor{Gray}
(8 dB, 2 dB) & 11\% & 6\% & 7\% \\
\hline
(6 dB, 2 dB) & 1\% & 2\% & 1\% \\
\hline
\end{tabular}
\end{table}

\section{Conclusion}\label{conclusion}

In this paper, we study the performance of a broadcast system using simultaneously time sharing and hierarchical modulation. With a particular rate policy, we present how to compute the rate for each receiver and illustrate the performance on two examples. The results show that the gain (in terms of throughput) of using hierarchical modulation can be significant, up to 17\%. It also points out the importance of the grouping strategy on the performance.

Future work will improve the comparison of the two schemes. An example with more receivers and a more accurate modeling of the satellite broadcast area will be considered. Also, the impact of the grouping strategies is worth studying in more detail. Finally, we would like to apply our idea to the DVB-S2 standard.

\appendix[Computation of the decoding thresholds]
The decoding thresholds are computed using the method described in \cite{wts}. First, for any modulation, the normalized capacity is defined by $\overline{C}_{mod} = \frac{1}{m} C_{mod}$, where  $C_{mod}$ is the modulation's capacity and m denotes the number of bits per symbol. Now we would like to obtain the decoding thresholds for the hierarchical 16-QAM with $\alpha=1$ at a target Bit Error Rate (BER) of $10^{-5}$. The HP and LP streams use both the 1/5-turbo codes. The method requires the performance curve (e.g., BER against $E_s/N_0$) for one reference modulation, which is the QPSK in the example. The decoding thresholds are computed as follow:
\begin{enumerate}

\item Use the performance curve of the QPSK with rate $R=1/5$ to get the operating point $\left(E_s/N_0\right)_{ref}$ corresponding to the desired performance. In the DVB-SH guidelines \cite[Table 7.5]{sh}, we find,
\begin{eqnarray*}
BER_{QPSK} \left( -3.9 \mbox{ dB} \right) = 10^{-5}.
\end{eqnarray*}

\item Compute the normalized capacity $\tilde R$ for the QPSK,
\begin{eqnarray*}
\tilde R_{HP} = \tilde R_{LP} = \overline{C}_{QPSK} \left( -3.9 \mbox{ dB} \right) \approx 0.2454,
\end{eqnarray*}

\item For the hierarchical 16-QAM, compute $E_s/N_0$ such as the normalized capacity at this SNR equals $\tilde R$. Then, the decoding threshold for the HP and LP streams are,
\begin{eqnarray*}
\left( E_s/N_0 \right)_{HP} &=& \overline{C}_{HP,\alpha=2}^{-1} \left( \tilde R_{HP} \right) = -2.46 \mbox{ dB}\\
\left( E_s/N_0 \right)_{LP} &=& \overline{C}_{LP,\alpha=2}^{-1} \left( \tilde R_{LP} \right) = 3.72 \mbox{ dB}
\end{eqnarray*}
\end{enumerate}



%



\nocite{*}
\bibliographystyle{IEEEtran}
\bibliography{biblio}

\end{document}